\documentclass[reprint, showpacs, superscriptaddress, amsmath, amssymb, aps, prl, twoside, nofootinbib]{revtex4-1}

\usepackage[normalem]{ulem}

\usepackage{graphicx}
\graphicspath{{figures/}}
\usepackage{bm}
\usepackage[version=3]{mhchem}
\usepackage{hyperref}
\usepackage{cleveref}
\usepackage{lipsum}
\usepackage{siunitx}
\usepackage{color}

\renewcommand{\thefootnote}{\fnsymbol{footnote}}
\newcommand\blfootnote[1]{%
  \begingroup
  \renewcommand\thefootnote{}\footnote{#1}%
  \addtocounter{footnote}{-1}%
  \endgroup
}

\begin{document}

\title{Coherent magneto-elastic domains in multiferroic \texorpdfstring{BiFeO$_3$}{BiFeO3} films}

\author{N. Waterfield Price}
\affiliation{\footnotesize Clarendon Laboratory, Department of Physics, University of Oxford, Parks Road, Oxford OX1 3PU, United Kingdom}
\affiliation{\footnotesize Diamond Light Source, Harwell Science and Innovation Campus, Didcot OX11 0DE, United Kingdom}

\author{R. D. Johnson}
\affiliation{\footnotesize Clarendon Laboratory, Department of Physics, University of Oxford, Parks Road, Oxford OX1 3PU, United Kingdom}
\affiliation{\footnotesize ISIS Facility, Rutherford Appleton Laboratory, Chilton, Didcot, OX11 0QX, United Kingdom}

\author{W. Saenrang}
\affiliation{\footnotesize Department of Materials Science and Engineering, University of Wisconsin-Madison, Madison, Wisconsin 53706, USA}

\author{F. Maccherozzi  }
\affiliation{\footnotesize Diamond Light Source, Harwell Science and Innovation Campus, Didcot OX11 0DE, United Kingdom}

\author{S. S. Dhesi}
\affiliation{\footnotesize Diamond Light Source, Harwell Science and Innovation Campus, Didcot OX11 0DE, United Kingdom}

\author{A. Bombardi}
\affiliation{\footnotesize Diamond Light Source, Harwell Science and Innovation Campus, Didcot OX11 0DE, United Kingdom}

\author{F. P. Chmiel}
\affiliation{\footnotesize Clarendon Laboratory, Department of Physics, University of Oxford, Parks Road, Oxford OX1 3PU, United Kingdom}

\author{C.-B. Eom}
\affiliation{\footnotesize Department of Materials Science and Engineering, University of Wisconsin-Madison, Madison, Wisconsin 53706, USA}

\author{P. G. Radaelli}\blfootnote{* p.g.radaelli@physics.ox.ac.uk}
\email{p.g.radaelli@physics.ox.ac.uk}
\affiliation{\footnotesize Clarendon Laboratory, Department of Physics, University of Oxford, Parks Road, Oxford OX1 3PU, United Kingdom}
\date{\today}

\begin{abstract}
The physical properties of epitaxial films can fundamentally differ from those of bulk single crystals even above the critical thickness. By a combination of non-resonant x-ray magnetic scattering, neutron diffraction and vector-mapped x-ray magnetic linear dichroism photoemission electron microscopy, we show that epitaxial $(111)$-\ce{BiFeO3} films support sub-micron antiferromagnetic domains, which are magneto-elastically coupled to a coherent crystallographic monoclinic twin structure. This unique texture, which is absent in bulk single crystals, should enable control of magnetism in \ce{BiFeO3} film devices via epitaxial strain.
\end{abstract}
\maketitle
\clearpage

Electrical manipulation of spins in insulators is a promising route to a new generation of fast, low consumption electronics \cite{Matsukura2015, Eerenstein2006, Ramesh2007}. Although direct electrical control of ferromagnets is challenging, much progress has been made towards electrical switching of \emph{antiferromagnetic} spins \cite{He2010}. Multiferroic \ce{BiFeO3} (BFO) is one of the most promising materials: at room temperature, BFO is both \emph{ferroelectric} and \emph{antiferromagnetic}, and its spins can be rotated by switching the direction of the electrical polarisation \cite{Zhao2006, Ratcliff2013}. Thus far, a fundamental limitation towards practical BFO devices has been the lack of understanding of the interplay between ferroelectricity, ferromagnetism and lattice distortions (ferroelasticity).  Using a combination of non-resonant x-ray magnetic scattering (NXMS), neutron diffraction and vector-mapped x-ray magnetic linear dichroism photoemission electron microscopy (XMLD-PEEM), we show that the antiferromagnetic domain structure of \SI{1}{\micro\metre} thick, epitaxial $(111)$--oriented BFO films  displays a $\approx$100 nm-scale texture, dramatically different from the mm-size features in bulk single crystals.  We also demonstrate that this magnetic texture is coherent (having matching topography and symmetry elements) with a pattern of monoclinic  domains at the nanoscale.  This texture is reminiscent of the dense polydomain states that are  thermodynamically stable in ferroelectric perovskites such as \ce{PbTiO3} in the presence of strain misfit \cite{Koukhar2001}.  This strongly suggests that the relaxed $(111)$--oriented BFO structure is not trigonal, but is a texture of coherent monoclinic micro twins. Besides providing a new pathway towards strain-engineering multiferroic domains in BFO, our approach yielded a detailed picture of the interplay between magnetism and lattice over 5 orders of magnitude in length scales, and could be applied to many classes of functional magnetic oxide devices.

\begin{figure}[h]
\includegraphics[width=0.5\textwidth]{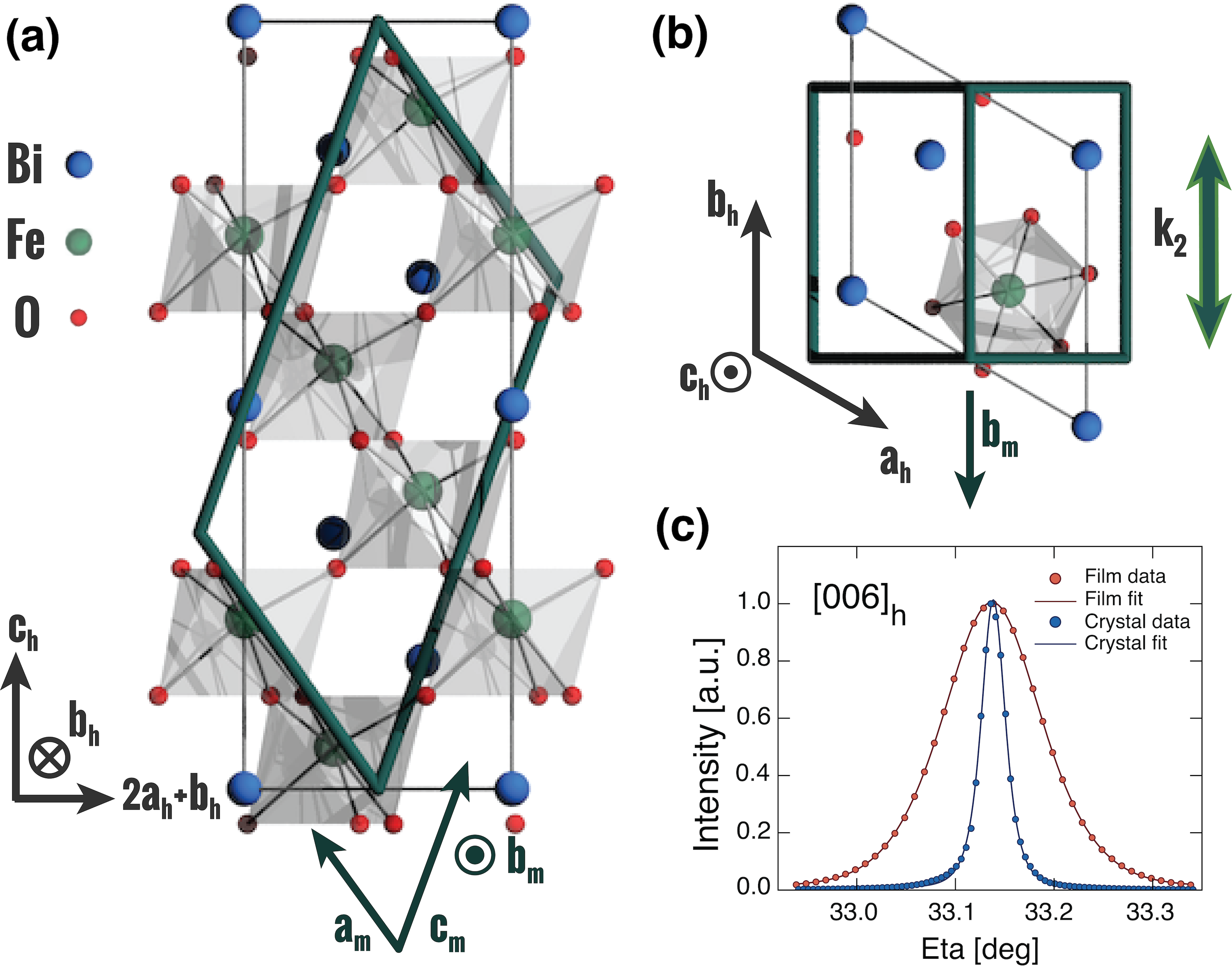}
\caption{\label{fig:structure} Monoclinic structure of \ce{BiFeO3}. \textbf{(a)} Side on (along $+\mathbf{b}_\mathrm{h}$ direction) and \textbf{(b)} top down (along $-\mathbf{c}_\mathrm{h}$ direction) projections of BFO crystal structure showing the relationship between the rhombohedral (black) and monoclinic (green) unit cells and their axes. The hexagonal and monoclinic settings are indicated by subscript h and m, respectively.  \textbf{(c)} Rocking curve scans of the $(006)_\mathrm{h}$ reflection in the film (red) and crystal (blue), respectively. The peak intensities have been normalised to unity.}
\vspace*{-0.7cm}
\end{figure}

Below its ferroelectric Curie temperature of $T_\mathrm{C} = \SI{1103}{\kelvin}$, bulk BFO is generally believed to possess a rhombohedrally-distorted perovskite structure with space group $R3c$ (pictured in \cref{fig:structure}(a) and (b)) \cite{Michel1969, Palewicz2007},  although very recent high-resolution synchrotron measurements have suggested a small monoclinic distortion \cite{Sosnowska2012}. The \ce{Bi^3+} and \ce{Fe^3+} cations are displaced away from their centrosymmetric positions along the $(111)$ (pseudo-cubic setting) axis \cite{Moreau1971}, producing a ferroelectric polarisation of $|\mathbf{P}| \sim \SI{100}{\micro\coulomb\per\centi\metre\squared}$, the largest of all known multiferroics. Below a N\'eel temperature of $T_\mathrm{N} \sim \SI{640}{\kelvin}$, bulk BFO orders antiferromagnetically. The structure can be described locally as G-type (i.e., with each spin being almost antiparallel to all its neighbours), but with a long-period ($l \sim \SI{62}{\nano\metre}$) cycloidal modulation with Fe magnetic moments rotating in a plane containing $\mathbf{P}$ and the magnetic propagation vector, $\mathbf{k}$ \cite{Sosnowska1982a}. Magnetic ordering breaks the symmetry of the three-fold axis, giving rise to three symmetry equivalent $\mathbf{{k}}$-domains, with magnetic propagation vectors $\mathbf{k_1} = (\delta,\delta,0)_\mathrm{h}$, $\mathbf{k_2} = (\delta,-2\delta,0)_\mathrm{h}$ and $\mathbf{k_3} = (-2\delta, \delta, 0)_\mathrm{h}$ where the subscript h denotes the hexagonal setting and $\delta = 0.0045$ at \SI{300}{\kelvin}. In high-quality bulk single crystals, antiferromagnetic cycloidal domains were previously imaged by NXMS and found to be of the order of a millimetre in size \cite{Johnson2013}.

Typical BFO device architectures are based on epitaxial thin films, and include a ferromagnetic layer, which can be switched via coupling to the BFO spins at the interface, as recently demonstrated for the $(001)$ film orientation \cite{Heron2014}.  In theory, $(111)$ oriented films should be the most attractive for applications due to the maximal out-of-plane polarisation and in-plane magnetic ordering, but are in practice plagued by electrical breakdown and fatigue problems \cite{Baek2011}, which  have been linked to a non-deterministic four-stage switching process \cite{Baek2011, Zou2012}.  Furthermore, unlike the $(001)$ case, in $(111)$ films three-fold symmetry is not broken by the substrate leaving the crystal structure reportedly rhombohedral \cite{Li2004, Bai2005, Kan2010, Ujimoto2012, Zavaliche2006, Das2006}. In this orientation, the magnetic modulation is not fixed by symmetry but can occur along three equivalent propagation directions. Therefore, precise electrical dialling of magnetism in $(111)$ BFO devices would require simultaneous control of the ferroelectric and ferroelastic switching pathways and of the magnetic modulation.

\begin{figure*}
\includegraphics[width=\textwidth]{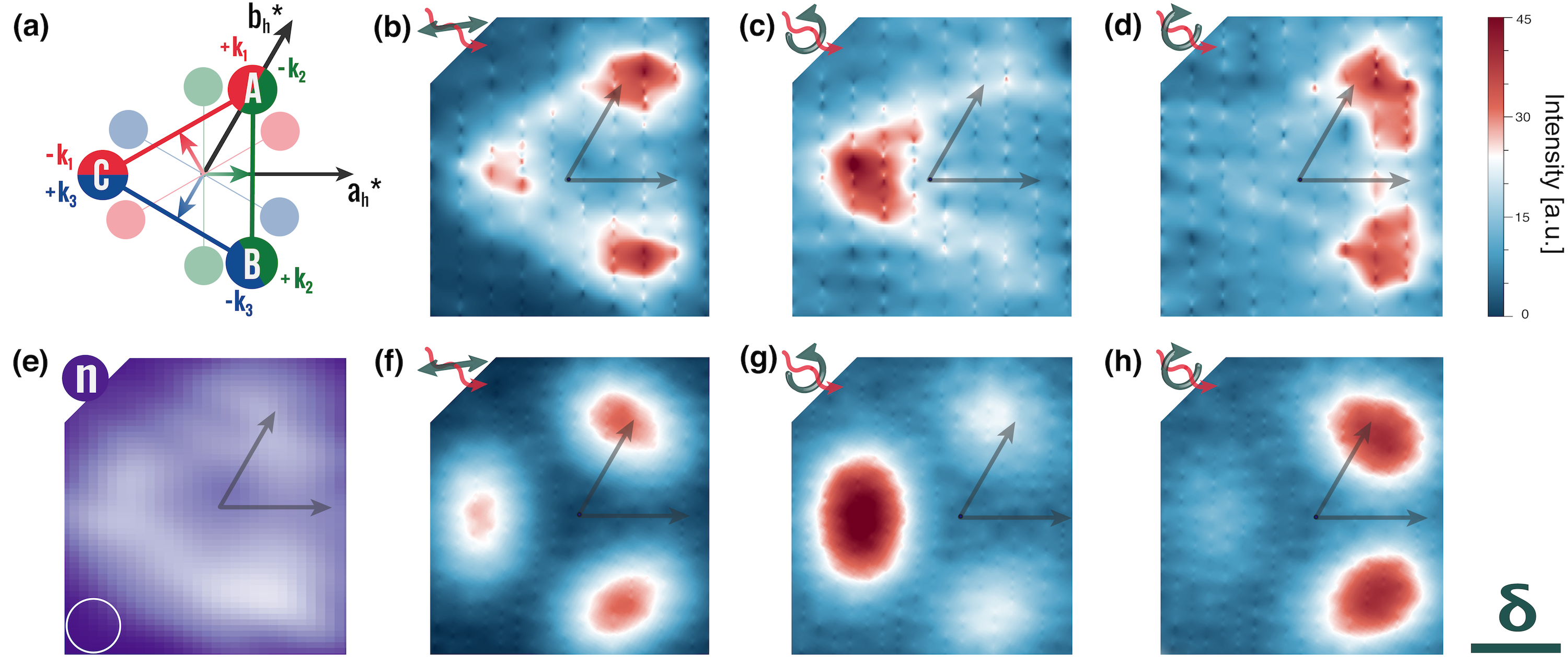}
\caption{\label{fig:magnetism} Magnetic diffraction reciprocal space maps. \textbf{(a)} Schematic of the monoclinic distortion model showing the undistorted (translucent) and distorted (opaque) diffraction patterns (see main text). \textbf{(b)--(d)} Reciprocal space maps about $(009)_\mathrm{h}$ measured at room temperature by NXMS with linear, left circular and right circular polarised light, respectively. The measured $(009)_\mathrm{h}$ position is slightly off center due to the sphere-of-confusion error of the diffractometer. \textbf{(e)} Reciprocal space map about the $(003)_\mathrm{h}$ measured by neutron diffraction at \SI{10}{\kelvin}. The white circle in the lower left corner indicates the instrumental resolution. \textbf{(f)--(h)} Simulated NXMS reciprocal space maps about $(009)_\mathrm{h}$ with linear, left circular and right circular polarised light, respectively. The scale bar in the bottom right shows the magnitude of the propagation vector (all images on the same scale) and the reciprocal lattice directions (in the hexagonal setting) are indicated by the black arrows.}
\vspace*{-0.6cm}
\end{figure*}

\SI{1}{\micro\metre} thick epitaxial films of $(111)$ BFO comprising a single ferroelectric domain were grown by double gun off-axis sputtering onto a $(111)$ surface normal \ce{SrTiO3} single crystal substrate \cite{Das2006} (see Supplemental Material S-I \cite{Suppl} \nocite{Eom1992,Vanderbilt2000,Xu2005,Xu2006,Liu2011,Wang2013b,Chapon2011,Blume1988} for further details). The films display excellent ferroelectric characteristics with a remnant polarisation along the $[111]$ direction of $\mathbf{P} = \SI{102}{\micro\coulomb/\cm^2}$ (see Supplemental Material S-I \cite{Suppl}). The full width at half maximum of the rocking curve of the $(006)_\mathrm{h}$ reflection (\cref{fig:structure}(c)) is \SI{0.12}{\degree}, which is comparably narrow in comparison to reported values for films grown by sputtering \cite{Das2006, Zavaliche2006}, pulsed laser deposition \cite{Ujimoto2012,Zavaliche2006,Chu2007} and molecular beam epitaxy \cite{ShuMin2014}, confirms the high crystalline quality of the films. This is however still four times as broad as for a bulk single crystal, indicating the presence of strain, inhomogeneity or mosaic spread.

To probe the magnetic structure of the film, we measured by NXMS and neutron diffraction at room temperature the magnetic satellite reflections due to the long-range incommensurate magnetic ordering, which occur near structurally-forbidden Bragg peaks such as the $\mathbf{N}=(009)_\mathrm{h}$ \cite{Johnson2013}. The NXMS experiments were performed on beamline I16 at Diamond Light Source (UK). The incident x-ray beam energy was tuned to \SI{4.9}{\kilo\electronvolt}, off-resonance of all chemical elements present in the sample, and the x-ray polarisation state was controlled using a \SI{100}{\micro\meter}-thick diamond phase-plate. The beam footprint at the sample was set to \SI{50}{\micro\metre} x \SI{90}{\micro\meter} rms. Single-crystal neutron diffraction measurements were performed on the WISH instrument at ISIS, the UK pulsed Neutron and Muon Spallation Source. The sample was oriented with the (00l)$_\mathrm{h}$ zone axis in backscattering geometry, which allowed for the highest resolution measurement of magnetic satellites of the (003)$_\mathrm{h}$ space-group-forbidden reflection. The neutron beam size at the sample position was \SI{20}{\milli\metre} x \SI{20}{\milli\meter}, fully illuminating the sample. Data taken at room-temperature and \SI{10}{\kelvin} were found to be entirely consistent but with an improved signal-to-background in the \SI{10}{\kelvin} dataset.

The diffraction pattern expected from each of the magnetic $\mathbf{k}$-domains is a pair of peaks at positions $\mathbf{N} \pm \mathbf{k}_i$, where $\mathbf{k}_i$ is the propagation vector of the corresponding magnetic domain, and this was indeed observed from a BFO single crystal with a small x-ray beam \cite{Johnson2013}. If populations of all three domains are illuminated, one should observe a star of six satellite peaks as depicted by the lightly-shaded circles in \cref{fig:magnetism}(a), with the red, green and blue peaks corresponding to propagation vectors $\mathbf{k}_1$, $\mathbf{k}_2$ and $\mathbf{k}_3$, respectively. Our experimental diffraction pattern (\cref{fig:magnetism}(b)) is in stark contrast with these expectations: within a large scanned volume of reciprocal space around $\mathbf{N}=(009)_\mathrm{h}$, we observe only \emph{three} reflections resembling a triangular shape --- a seemingly impossible situation, since for any satellite at $\mathbf{N} + \mathbf{k}$ there \emph{must} be a corresponding satellite at $\mathbf{N} - \mathbf{k}$ as the Fourier decomposition of a real, periodic magnetic structure necessarily contains $+\mathbf{k}$ and $-\mathbf{k}$ components. The only explanation is that each of the peaks in the triangle contains contributions from two domains (thereby preserving a total of six satellites), with the midpoint of each pair of satellites being displaced away from the nominal $(009)_\mathrm{h}$\ position. In fact, the observed diffraction pattern is well modelled by introducing a small monoclinic distortion which translates the six magnetic satellites in reciprocal space such that pairs of neighbouring reflections lie approximately on top of each other, resulting in three composite peaks $\{+\mathbf{k}_\mathrm{1}, -\mathbf{k}_\mathrm{2}\}$, $\{+\mathbf{k}_\mathrm{2}, -\mathbf{k}_\mathrm{3}\}$ and $\{+\mathbf{k}_\mathrm{3}, -\mathbf{k}_\mathrm{1}\}$ (\cref{fig:magnetism}(a)). The best fit to the experimental patterns is achieved by tilting the $c^*_\mathrm{h}$ axis by $\SI{0.030}{\degree} \pm \SI{0.005}{\degree}$ away from the $[001]_\mathrm{h}$ direction and \emph{orthogonal} to $\pm \mathbf{k}_i$. In real space, this corresponds to changing both the monoclinic angle $\beta_\mathrm{m}$ and the $c_\mathrm{m}/a_\mathrm{m}$ ratio (\cref{fig:structure}(c)), whilst fixing the direction of $[101]_\mathrm{m}$ to the surface normal. Our monoclinic structure (\cref{fig:structure}(a), (b)) has the same symmetry as the M$_\mathrm{A}$ or M$_\mathrm{B}$ fully relaxed structures, of $(100)$ and $(110)$ films of similar thickness \cite{Kan2010}. More details on the crystallography are given in the Supplemental Material (S-II \cite{Suppl}). The observed peak intensity is in quantitative agreement with theoretical calculations for three cycloidal domains (simulated map in \cref{fig:magnetism}(f), see Supplemental Material S-V \cite{Suppl} for full details).  For the experimental geometry used to collect the data in \cref{fig:magnetism}(b) ($\sigma$ linear polarisation, beam polarised perpendicular to the diffraction plane), the intensity calculation yields
\begin{equation}
  \frac{I_\mathrm{C}^\sigma}{I_{\mathrm{A}, \mathrm{B}}^\sigma}  \approx \frac{\sqrt{3}}{2},
\end{equation}
where the subscript and superscript denote the peak (see \cref{fig:magnetism}(a)) and incident x-ray polarisation, respectively. This agrees excellently with the observed diffraction pattern. To further test this model, we employed circularly polarised x-rays, since the intensities of the magnetic satellites are highly sensitive to the left/right photon helicity and the relative direction of the magnetic propagation vectors. A similar calculation to the linear case yields a peak intensity ratio of
\begin{equation}
  \frac{I_\mathrm{C}^\gamma}{I_{\mathrm{A}, \mathrm{B}}^\gamma} \approx \frac{1 + \gamma}{1 - \gamma/2}
\end{equation}
where $\gamma$ = 1(-1) for left-(right-)circularly polarised light, which reproduces the experimental polarisation dependence well (\cref{fig:magnetism}(c), (d)) as can be seen in the simulation (\cref{fig:magnetism}(g), (h)). In principle, the monoclinic distortion should also be observable as a splitting of the charge peaks. However, since the width of the charge peaks is an order of magnitude larger than the expected split, we could not observe this effect even at high Miller indices.

The following essential features of our model are worth emphasising: firstly, the intensities of the three peaks in \cref{fig:magnetism}(b) are observed uniformly over the surface of the film and consistent with scattering from equal magnetic (and hence monoclinic) domain populations. This implies that the characteristic domain size must be significantly smaller than the beam profile of \SI{50}{\micro\metre} x \SI{90}{\micro\meter},  placing an upper limit of the order of a few microns on the domain size, in stark contrast to mm-size domains for the bulk single crystals \cite{Johnson2013}. Secondly, the magnetic and structural (monoclinic) domains must be \emph{topographically coherent} with each other -- in other words, monoclinic domain boundaries are also magnetic domain boundaries. This constraint is naturally imposed by the fact that the symmetry of both the structural and magnetic domains is $Cc$, which is hence the magneto-structural domain symmetry, a fact that also holds for the large single crystal domains. Moreover, the direction of the $c^*_\mathrm{h}$ axis tilt \emph{must} be the same in each magneto-structural domain type, otherwise a double rather than a single triangle would be observed. In fact, the same scattering is observed in our neutron diffraction data (\cref{fig:magnetism}(e)) collected with the beam illuminating the entire sample, strongly suggesting that the whole film comprises a coherent twin pattern of monoclinic micro-domains (see Supplemental Material S-III \cite{Suppl} for further details). Although the three composite peaks are clearly visible in \cref{fig:magnetism}(e), the neutron data are rather broad and display finite intensity in the intermediate regions between the three composite peaks, indicating that over the whole sample there may be disorder in the direction of the propagation vector.

\begin{figure}
\includegraphics[width=0.5\textwidth]{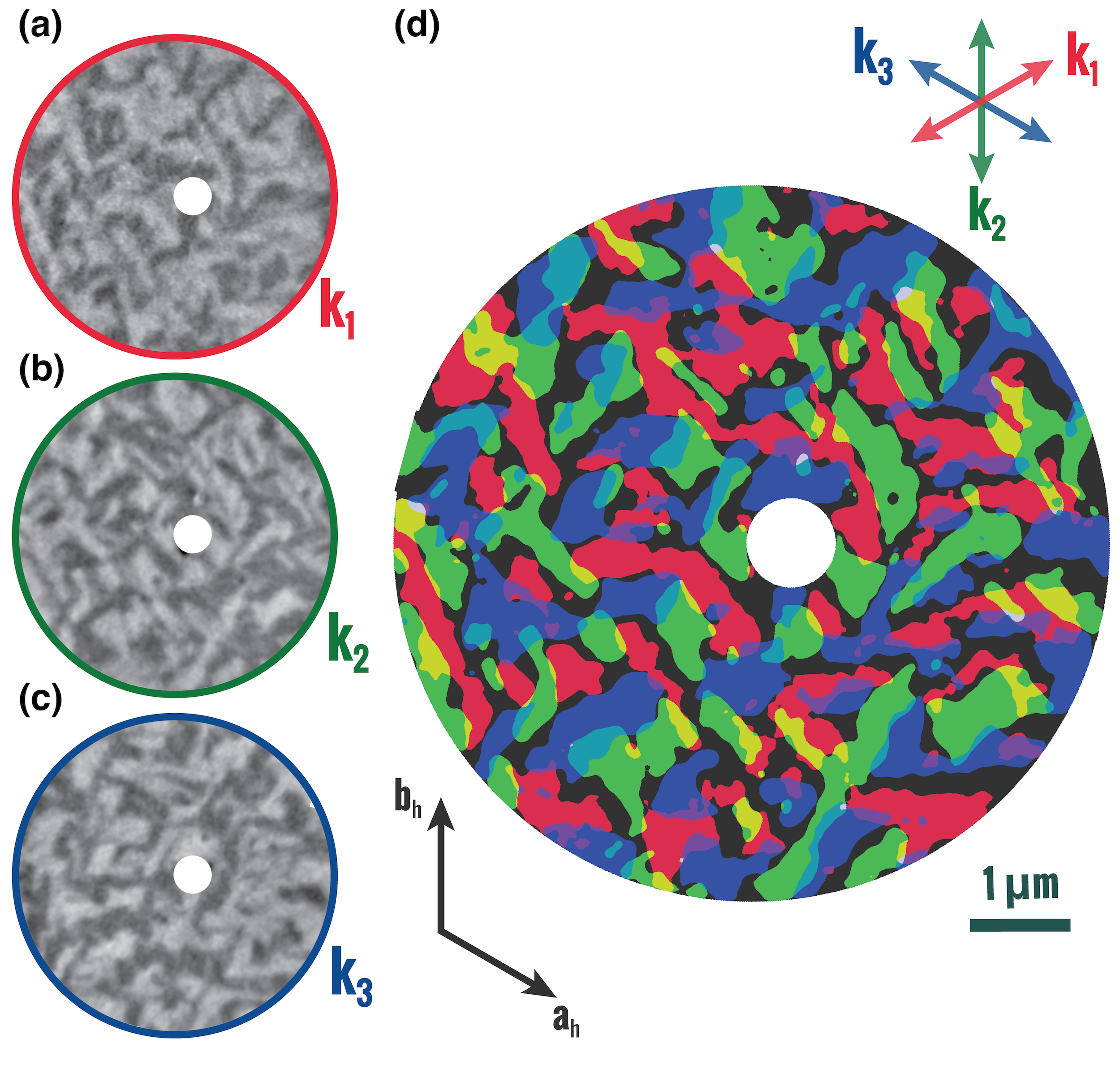}

\vspace*{-0.3cm}\caption{\label{fig:peem} PEEM imaging of magnetic $\mathbf{k}$-domains in $[111]$-\ce{BiFeO3}-films. \textbf{(a)--(c)} PEEM images obtained using x-ray linear dichroism (XMLD) with the light areas corresponding to domains of $\mathbf{k}_1$, $\mathbf{k}_2$ and $\mathbf{k}_3$ propagation vectors, respectively. \textbf{(d)} Discrete vector map of the $\mathbf{k}$-domains obtained from (a)--(c), where the direction of the propagation vector is indicated by the coloured arrows. The scale bar refers to (d).}
\vspace*{-0.7cm}
\end{figure}

We have directly imaged the antiferromagnetic cycloidal micro-domains in our $(111)$ BFO film by angle-resolved XMLD-PEEM. Measurements were carried out on beam line I06, Diamond light Source, UK using an Elmitec SPELEMM-III microscope. Linearly polarised x-rays at a grazing incidence angle of \ang{16} were used to record images at energies E1 = \SI{708.2}{\electronvolt} and E2 = \SI{708.9}{\electronvolt}, corresponding to \SI{-0.2}{\electronvolt} and \SI{+0.4}{\electronvolt} from the Fe L$_{3}$ absorption edge, respectively. XMLD asymmetry, $(I_{E1} - I_{E2})/(I_{E1} + I_{E2})$, was calculated at each pixel.

Contrast features with length scales of the order of \SI{1}{\micro\metre} are clearly visible in all raw images collected with the polarisation of the light in the plane of the film (\cref{fig:peem}(a)--(c)). The contrast is observed to be much weaker when the polarisation is out of plane (a weak contrast is expected due to the ~\SI{16}{\degree} grazing incidence angle of the x-rays).  This is consistent with cycloidal magnetic domains with in-plane propagation vectors and planes of rotation orthogonal to the film plane. We note that any ferroelastic contribution to the PEEM contrast \cite{Zhao2006} can be ruled out as our (111) film is a single ferroelectric \emph{monodomain}. The magnetic origin of the PEEM contrast is further corroborated by its temperature dependence (\SI{60}{\percent} reduction at \SI{475}{\kelvin}), which was found to be quantitatively consistent with the temperature dependence of the ordered moment \cite{Ramazanoglu2011}, and was reversed upon cooling.

The intensity of the XMLD signal in a single cycloidal magnetic domain, obtained by a spatial average of the collinear antiferromagnetic case \cite{Alders1995}, is proportional to $ \mathbf{M}^2 \left(3/2 \cos^2(\hat{\mathbf{k}} \cdot \hat{\mathbf{E}}) +3/2 \cos^2(\hat{\mathbf{z}} \cdot \hat{\mathbf{E}})- 1\right)$, where $\mathbf{M}^2$ is the square of the local magnetic moment, while $\hat{\mathbf{k}}$ and $\hat{\mathbf{z}}$ are the directions of the propagation vector and of the surface normal, respectively, and $\hat{\mathbf{E}}$ is the polarisation vector of the light. The signal is maximum ($\propto + \mathbf{M}^2/2$) for $\hat{\mathbf{E}}$ along the propagation vector, in which case the signal from the other two domains is $\propto - 5\mathbf{M}^2/8$. When $\hat{\mathbf{E}}$ is perpendicular to the film surface, the signal is the same for all three domains ($\propto + \mathbf{M}^2/2$), consistent with our observations of weak contrast.  A full spatial map of the cycloidal domains (\cref{fig:peem}(d)) can be constructed by appropriately combining images collected at multiple sample rotations (see Supplemental Material S-IV \cite{Suppl})  \cite{Moya2012}. The contrast in \cref{fig:peem} is consistent with the sample surface consisting of approximately equal populations of the three $\mathbf{k}$-domains. Furthermore, the orientations of the domain boundaries mostly correspond to those predicted by the twinning model inferred from NXMS (see Supplemental Material S-III \cite{Suppl}).

The remarkable texture we observe can most naturally be explained assuming a slight monoclinic distortion in our relaxed $(111)$ films of BFO, consistent with the unit cell of $(001)$ and $(110)$ relaxed films \cite{Kan2010} and single crystals \cite{Sosnowska2012}. Unlike in the $(100)$ and $(110)$ oriented films, the misfit strain and symmetry mismatch from the three-fold symmetric $(111)$ \ce{SrTiO3} substrate would stabilise a sub-micron texture of coherent monoclinic twins, to which the magnetic propagation vector becomes locked. Our observations clearly demonstrate that multiferroic domains in $(111)$ BFO films break three-fold symmetry in \emph{both} magnetic \emph{and} crystal sectors, and should therefore be controllable by strain and substrate miscut as for other BFO orientations. By appropriate substrate choice, biasing the growth of a single monoclinic domain with a non three-fold symmetric substrate, one should be able to grow $(111)$ BFO films as a single multiferroic domain, combining the full, surface normal electrical polarisation with a coherent magnetic structure, which may provide exceptionally strong magnetic interface coupling and possibly even a pre-biassed unique ferroelectric switching path. More broadly, the combination of NXMS, neutron diffraction and vector-mapped XMLD-PEEM yielded a detailed picture of the interplay between magnetism and lattice from $\SI{1}{\centi\metre}$ to less than $\SI{100}{\nano\metre}$ --- an approach applicable to many other functional antiferromagnets in device configurations.
\newpage
The authors would like to thank Pascal Manuel for assistance with the neutron diffraction measurements and Jungwoo Lee for assistance with the in-plane reciprocal space map measurement. We would also like to thank Laurent Chapon for the use of his x-ray diffraction simulation code. We acknowledge Diamond Light Source for time on Beamline I16 under Proposal MT10201-1 and time on Beamline I06 under Proposal NT12336. The work done at the University of Oxford (N.W.P., R.D.J., F.P.C. and P.G.R.) was funded by EPSRC Grants EP/J003557/1, entitled ``New Concepts in Multiferroics and Magnetoelectrics'' and EP/M020517/1, entitled ``Oxford Quantum Materials Platform Grant''. The work at University of Wisconsin-Madison (W.S. and C.-B. E.) is supported by the Army Research Office under Grant No. W911NF-13-1-0486.

In accordance with the EPSRC policy framework on research data, access to the data will be made available from \cite{Data}.

\bibliography{bfo_references}

\end{document}